\documentclass[aps,epsf,preprint,showpacs]{revtex4}
\usepackage{amsmath}
\usepackage{epsfig}

\begin{document}

\title{Review of recent developments in the random--field Ising model}

\author{Nikolaos G. Fytas$^{1}$, V\'{i}ctor Mart\'{i}n-Mayor$^{2,3}$, Marco Picco$^4$, and Nicolas Sourlas$^5$}

\affiliation{$^1$Applied Mathematics Research Centre, Coventry
University, Coventry CV1 5FB, United Kingdom}

\affiliation{$^2$Departamento de F\'isica T\'eorica I, Universidad
Complutense, 28040 Madrid, Spain}

\affiliation{$^3$Instituto de Biocomputac\'ion y F\'isica de
Sistemas Complejos (BIFI), 50009 Zaragoza, Spain}

\affiliation{$^4$Sorbonne Universit\'es, Universit\'e Pierre et
Marie Curie - Paris VI,
Laboratoire de Physique Th\'eorique et des Hautes Energies, CNRS UMR 7589, \\
4, Place Jussieu, 
75252 Paris Cedex 05, France}

\affiliation{$^5$Laboratoire de Physique Th\'eorique de l'ENS,
\'Ecole Normale Sup\'erieure,PSL Research University, Sorbonne
Universit\'es, UPMC Univ. Paris 06, CNRS, 75005 Paris, France}

\date{\today}

\begin{abstract}
A lot of progress has been made recently in our understanding of
the random--field Ising model thanks to large--scale numerical
simulations. In particular, it has been shown that, contrary to
previous statements: the critical exponents for different
probability distributions of the random fields and for diluted
antiferromagnets in a field are the same. Therefore, critical
universality, which is a perturbative renormalization--group
prediction, holds beyond the validity regime of perturbation
theory. Most notably, dimensional reduction is restored at five
dimensions, \emph{i.e.}, the exponents of the random--field Ising
model at five dimensions and those of the pure Ising ferromagnet
at three dimensions are the same.
\end{abstract}


\maketitle

\section{Introduction}
\label{sec:intro}

The random--field Ising model (RFIM) is the simplest disordered
system. Its Hamiltonian is
\begin{equation}
\mathcal{H} = - J \sum_{\langle xy \rangle} \sigma_x \sigma_y -
\sum_{x} h_x \sigma_x,
\end{equation}
where $\sigma_x = \pm 1$ on a hypercubic lattice in $D$ dimensions
with nearest--neighbor ferromagnetic interactions. The $h_{x}$ are
independent random magnetic fields obeying different probability
distributions. The RFIM has a long history. Let us remind some of
the most important results.

The perturbative renormalization group (PRG) can be carried out to
all orders of perturbation theory (in fact this is the only known
case where PRG can be carried to all orders). It predicts
dimensional reduction and supersymmetry~\cite{Aharony,Y,PS1}. This
means that the critical exponents of the RFIM in $d$ dimensions
are the same as the exponents of the pure Ising model in $d-2$
dimensions. It has been proven that dimensional reduction is not
true in three dimensions~\cite{BK}.

There are some other open questions we would like to address in
this paper. What about other predictions of PRG? Are some
predictions still true? What about higher dimensions? Why PRG
breaks down at three dimensions? There have been recently very
large--scale simulations which clarify these questions. The main
conclusions are: contrary to previous statements universality is
true in three, four, and five dimensions. Diluted antiferromagnets
in a field are in the same universality class with the RFIM. PRG
breaking is a low--dimensional phenomenon. PRG and dimensional
reduction are restored at five dimensions. There is a maximum
violation of self--averaging in the distribution of low lying
excited states near the critical point.

\section{Universality}
\label{sec:universality}

The explanation of critical universality is a major success of the
renormalization group. In the case of the RFIM, PRG predicts that
different random--field Ising models, where the random fields are
drawn from different probability distributions, belong to the same
universality classes. Also, more surprisingly, diluted
antiferromagnets in a field are predicted to belong to the same
universality class. These two predictions have been recently shown
numerically to be true in three, four, and five
dimensions~\cite{FMMa,FMMb,PiS,FMMPSa,FMMPSa2,FMMPSb}, despite the
failure of PRG. Previous numerical simulations are incompatible
with these predictions for universality.

Because of these older simulations, the prevailing view was that
universality is not valid for the RFIM because of the failure of
PRG. This view has changed thanks to recent simulations by Fytas
and Mart\'{i}n-Mayor~\cite{FMMa,FMMb}. They considered double
Gaussian (considering also the bimodal and Gaussian limits) and
Poissonian random--field distributions in three dimensions. They
found the same exponents for all distributions of the random
fields.

What explains the disagreement with previous work?
\emph{Subdominant corrections to scaling}. For finite sizes there
is no simple power--law behavior near the critical point. There
are subdominant corrections to single power--law behavior. Large
lattice sizes and high--precision data are needed in order to
measure the subdominant corrections to scaling. Fytas and
Mart\'{i}n-Mayor simulated systems with linear sizes up to $L =
196$ and $10^7$ samples per size. Furthermore you need the
appropriate theoretical framework in order to analyze these
corrections.

In field theory, subdominant corrections are controlled by the
Callan--Symanzik $\beta(g)$ function. $ \omega = \left.\frac{\mathrm{d}\beta}{\mathrm{d}g}\right|_{g=g^{\ast}}$ is a universal exponent and controls the
leading corrections to scaling. The same exponent, $\omega$,
controls corrections to scaling for all observables. When you have
good enough data allowing you to compute those non--leading
corrections, you find that all considered probability
distributions of the random fields have the same exponents, thus
confirming universality.

Subsequently Picco and Sourlas~\cite{PiS}, using the same methods,
have shown that the three-dimensional diluted antiferromagnets in
a field belong to the same universality class with the RFIM,
contrary to previous assertions. Fytas \emph{et al.}, computed
critical exponents at four and five dimensions for both the
Gaussian and Poissonian distributions. Exponents are the same for
both distributions and universality is valid also at four and five
dimensions, see Refs.~\cite{FMMPSa,FMMPSa2,FMMPSb}. Universality
is of course not only valid in PRG but is a general property of
the renormalization group, but it is in general very hard to show
that two different physical systems belong to the same
universality class except by using the PRG. This is particularly
the case of the RFIM.

\begin{figure}[htbp]
\includegraphics*[width=11 cm]{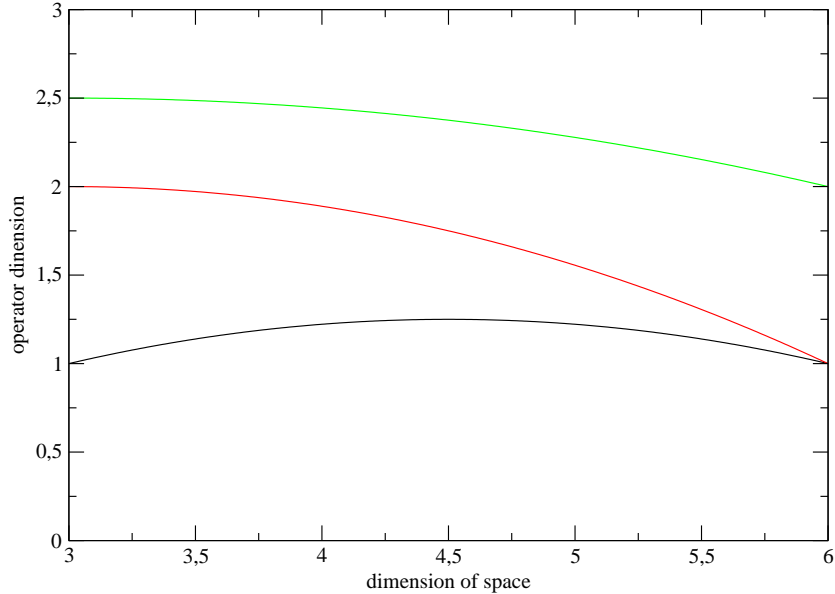}
\caption{\label{fig:operator} (color online) Pictorial
representation of the non--crossing mechanism that protects
universality (see relevant discussion in the main text). The
figure, which has an illustrative purpose only, depicts the
operator dimension of three generic operators as a function of the
dimension of space.}
\end{figure}

Why critical universality, as established by the PRG is valid in
three dimensions while PRG is broken? In PRG, universality is
established using Wilson's operator product expansion and the
$\epsilon$ expansion, \emph{i.e.} close to the upper critical
dimension, $D_{\rm u} = 6$  in the case of the RFIM. One
classifies the operators into relevant, marginal, or irrelevant
according to their scaling dimensions. The scaling dimensions of
the operators $D_{O}$ are functions of the dimension of space $d$.
The $D_{O}(d)$'s change when the dimension of space is changed. A
necessary and sufficient condition for non changing universality
classes as the dimension of space $d$ varies is for the scaling
dimensions $D_O(d)$ of the leading operators not to cross when the
dimension of space is lowered from $d = D_{\rm u}$ down to $d = 3$
as this is illustrated in Fig.~\ref{fig:operator}.

If this is the case, the classification of operators into
relevant, marginal, and irrelevant remains unchanged when the
dimension of space is lowered.

The $\epsilon$ expansion computes the dimension of the operators
and their derivatives with respect to $d$ at $d = D_{\rm u}$. Why
the classification is it still valid for $\epsilon=3$ when PRG is
broken? In fact we do not know of \emph{any case}  of the
classification into universality classes using the $\epsilon$
expansion not been valid at lower dimensions. Why is this so?

The reason for the scaling dimensions of operators not to cross
when the dimension of space decreases is probably the
following~\cite{Sourlas}. Scaling dimensions of operators are
eigenvalues of the scaling transformations, \emph{i.e.} of the
group of dilatations of space. It is well known from the early
days of quantum mechanics that in the generic case the eigenvalues
of operators, that is the eigenvalues of the matrices in their
matrix representation, do not cross if one changes a single
parameter~\cite{LL}. This phenomenon is called repulsion of
eigenvalues. If universality is valid around $d = 6$ dimensions
there is big chance to be valid at $d = 3$ dimensions because of
eigenvalue repulsion. Validity of universality at three dimensions
does not depend on the validity of PRG at this dimension. The
previous argument can be inverted. If it is found by other means,
like experiments or numerical simulations, that universality
classes in three dimensions coincide with those established by the
$\epsilon$ expansion, it means that PRG is valid near the upper
critical dimension $D_{\rm u}$, \emph{i.e.} the epsilon expansion
is valid.

\section{Discussion on the dimensional reduction restoration}

\begin{figure}[htbp]
\includegraphics*[width=13 cm]{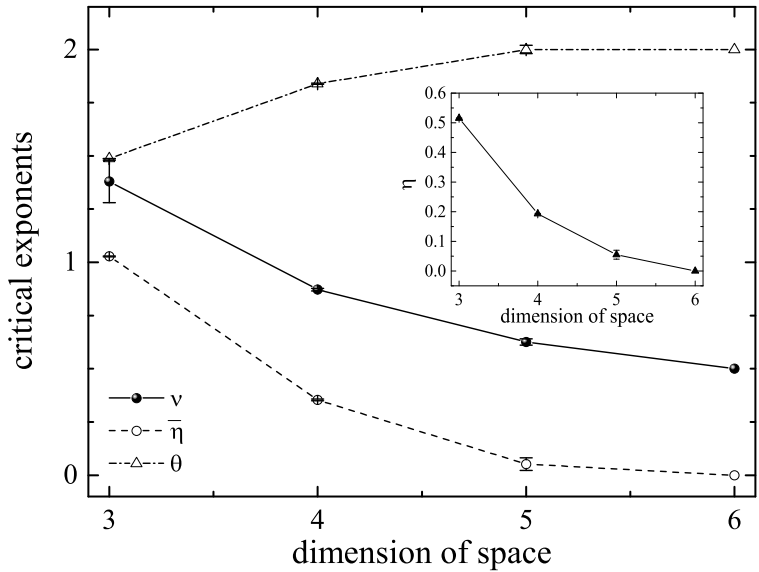}
\caption{\label{fig:exponents} Numerical estimates of critical
exponents as a function of the dimension of space for the
$3d$~\cite{FMMa}, $4d$~\cite{FMMPSa,FMMPSa2}, and $5d$
RFIM~\cite{FMMPSb}. The data points corresponding to $d = 6$ are
the mean--field exponents.}
\end{figure}

Recent numerical simulations computed the critical exponents for
the Gaussian and Poissonian probability distributions of the
random fields at four~\cite{FMMPSa,FMMPSa2} and five
dimensions~\cite{FMMPSb}. It was found that PRG and dimensional
reduction is violated at four dimensions. On the contrary our
simulations are compatible with the validity of PRG and
dimensional reduction in five dimensions. In five dimensions we
simulated Gaussian and Poissonian random fields, for linear sizes
$4\le L \le 28 $ and $10^7$ samples per size. Numerical results
for our estimates of critical exponents as a function of the space
dimensionality are summarized in Fig.~\ref{fig:exponents}. In the
main panel we show the cases for $\nu$, $\overline{\eta}$, and the
violation of hyperscaling exponent $\theta =
2-\overline{\eta}+\eta$~\cite{FMMPSa2}. In the corresponding inset
we present the anomalous dimension $\eta$ on its own for clarity
reasons.

The most compelling evidence comes from fitting the data
\emph{assuming} dimensional reduction, \emph{i.e.} assuming that
the critical exponents of the RFIM in five dimensions are equal to
those of the pure Ising model in three dimensions, which are known
very precisely. As all the exponents are assumed to be known there
are very few parameters to fit. The quality of the fit, $\chi^2$
per degree of freedom, is excellent: $3.43/7$ for $\nu$,
$13.37/11$ for $\eta $, and $4.15/7$ for the difference $2 \eta -
\overline{\eta}$. We conclude that PRG and dimensional reduction
are restored at five dimensions and that the breaking of PRG is a
low--dimensional phenomenon. Tarjus \emph{et
al.}~\cite{Tar1,Tar2,Tar3} using functional renormalization group,
also argue that dimensional reduction is restored for dimensions
larger than $5.1$.

Why PRG is valid at higher dimensions? What is the reason of PRG
failure at low dimensions? As PRG is valid at all orders of
perturbation theory, the reason of its breaking must be non
perturbative. Parisi and Sourlas~\cite{PS0} proposed that the
reason is the formation of bound states, which is a non
perturbative phenomenon. The mass of the bound state provides a
new length scale which is not taken into account in the
traditional PRG analysis.

The argument for the formation of bound states is the following:
first Kardar \emph{et al.}~\cite{K1,K2} observed that interactions
among replicas are attractive -- this is not the case for branched
polymers and dimensional reduction holds in that case. For the
random interface problem they computed the spectrum of the
effective Hamiltonian with Bethe ansatz and found the lowest
states to be bound states. Then Br\'ezin and De
Dominicis~\cite{BDD1,BDD2} wrote the Bethe--Salpeter equation for
$\langle s^\alpha s^\beta (0) s^\alpha s^\beta (x) \rangle $ in
the RFIM, where $\alpha$ and $\beta$ are two different replica
indices and they found an instability in the Bethe--Salpeter
kernel for $d < 6$ most probably implying the existence of bound
states. As remarked by Parisi~\cite{Pa83},
\begin{equation}
\label{eq:2} \langle s^{\alpha}(0) s^{\alpha}(x) \rangle =
\overline{\langle \sigma(0)\sigma(x)\rangle},
\end{equation}
where $\overline{\langle \sigma(0)\sigma(x)\rangle}$ is the
average over samples of $\langle \sigma(0)\sigma(x)\rangle$, and
\begin{equation}
\label{eq:3} \langle s^{\alpha} s^{\beta} (0) s^{\alpha} s^{\beta}
(x) \rangle = \overline {(\langle \sigma(0) \sigma(x)\rangle)^2},
\end{equation}
where the replica indices $ \alpha \ne \beta$.

\begin{figure}
\begin{center}
{\epsfxsize=8cm\epsffile{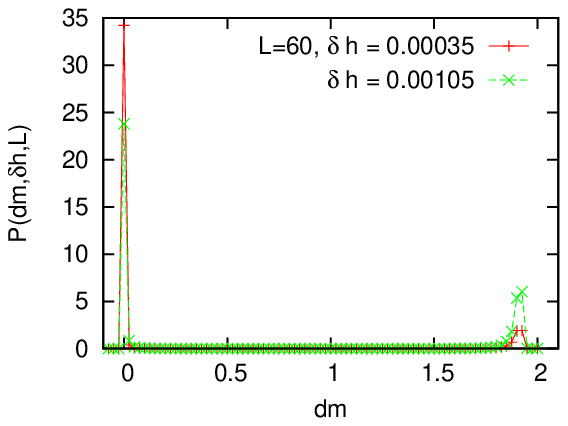}}
{\epsfxsize=8cm\epsffile{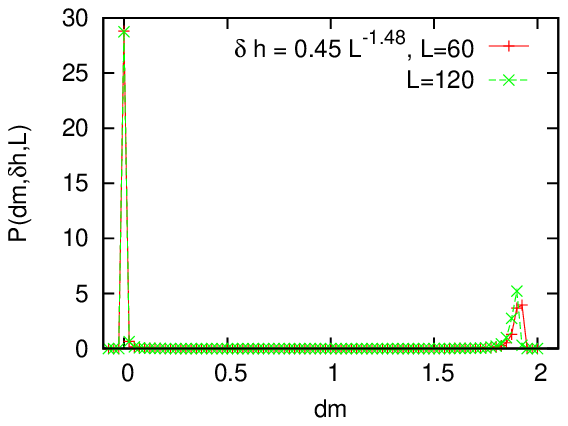}}
\end{center}
\caption{(Color online) Illustration of the probability
distribution $P(dm, dj, \delta h,L)$ for $dj = 0$. Left panel: $L
= 60$ and two values of $\delta h$. Right panel: $L=60$, $L=120$
and $\delta h =0.45 L^{-\gamma /\nu}$. \label{fig:dm1}}
\end{figure}

Parisi and Sourlas~\cite{PS0} found that for large $x$ in three
dimensions
\begin{equation}
\label{eq:4a} \overline{\langle \sigma(0)\sigma(x)\rangle} \sim
\exp{(-mx)}
\end{equation}
and
\begin{equation}
\label{eq:4b} \overline{(\langle \sigma(0)\sigma(x)\rangle)^2 }
\sim \exp{(-mx)}
\end{equation}
with the \emph{same} $m$! $m$ is the mass of the intermediate
state with the lowest mass. This means that there exists a state
in replica field theory that couples to both the $s^{\alpha} (x) $
and $s^{\alpha} s^{\beta} (x)$ operators. Not such state exists in
perturbation theory. They concluded to the existence of a new
state, not present in perturbation theory, which couples to both
$s^{\alpha}(x)$ and $s^{\alpha}s^{\beta}(x)$. This must be a bound
state, in agreement with Br\'ezin and De Dominicis. They also
pointed out that the space dimension plays a crucial role in the
formation of bound states. In the formation of bound states there
is a competition between the attractive forces  and the size of
the available phase space. The size of  phase space increases with
the dimension of space making the formation of bound states more
difficult in higher dimensions. We expect that for high enough
dimensions bound states will no longer exist, and PRG predictions
should hold. This argument does not predict at which dimension PRG
is restored.

Another remarkable fact is the maximum violation of
self--averaging in the sample to sample fluctuations, as explained
below. In order to compute the magnetic susceptibility, one can
add a small translation invariant magnetic field $\delta h$. For
each sample compute $m_0$, the ground--state magnetization at
$J=J_{\rm c}+dj$, where $J_{\rm c}$ denotes the critical point,
add $\delta h$ and find the new ground--state magnetization $m'$.
The resulting change of the ground--state magnetization due to
$\delta h$ is $dm = m_0-m'$.

\begin{figure}
\begin{center}
{\epsfxsize=8cm\epsffile{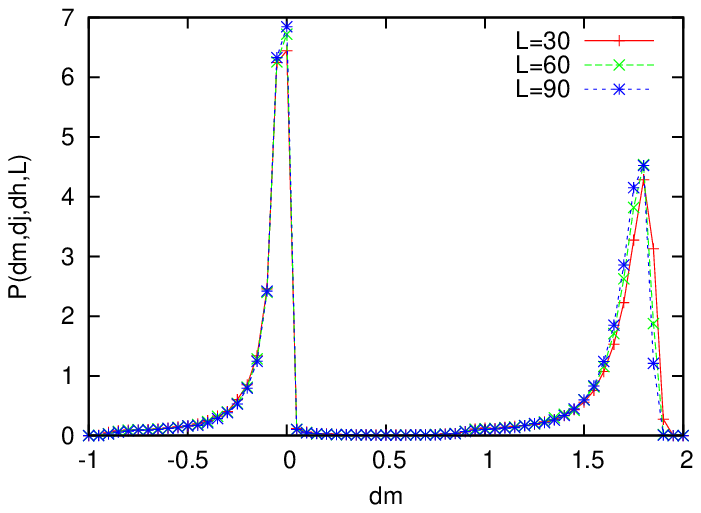}}
{\epsfxsize=8cm\epsffile{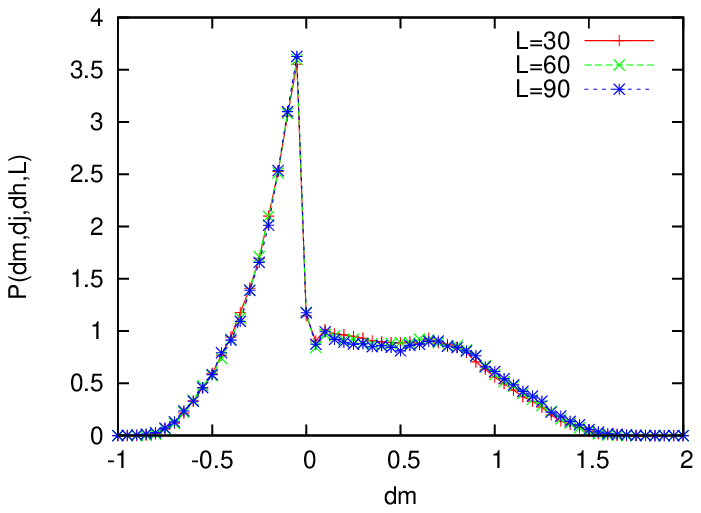}}
\end{center}
\caption{(Color online) Illustration of the probability
distribution $P(dm, dj, \delta h,L)$ for $dj=0.3 L^{- 1/\nu }$,
$\delta h = 2.2 L ^{- \gamma /\nu}$ (left panel) and $dj=0.6 L^{-
1/\nu}$, $\delta h = 1.1 L ^{-\gamma/\nu}$ (right panel).
\label{fig:dm2} }
\end{figure}

Consider the probability distribution $P(dm, dj, \delta h,L)$ of
$dm$ over the random--field samples, see Fig.~\ref{fig:dm1}. The
left panel of the figure shows $P(dm, dj, \delta h,L)$ of $dm$ for
$dj = 0$, \emph{i.e.} at $J=J_{\rm c}$, for $L=60$ and two values
of $\delta h$. We see it is a bimodal probability distribution.
Increasing $\delta h$ only modifies the size of the two peaks.
Note the bound $0\leq dm\leq 2$, implying that our bimodal
distribution, with one of the peaks at $dm\sim 2$, is a maximum
violation of self--averaging. The right panel of
Fig.~\ref{fig:dm1} illustrates $P(dm, dj, \delta h,L)$ of $dm$ for
$dj=0$, \emph{i.e.} at $J=J_{\rm c}$ for $L=60$ and $L=120$ and
$\delta h =0.45 L ^{-\gamma /\nu}$. $\gamma$ is the critical
exponent of the magnetic susceptibility. Clearly, $P(dm, dj,
\delta h,L)$ obeys finite--size scaling.

The next two panels in Fig.~\ref{fig:dm2} present $P(dm, dj,
\delta h,L)$ for $dj = 0.3 L^{-1/\nu}$ and $\delta h = 2.2 L
^{-\gamma /\nu}$, and $dj = 0.6 L^{-1/\nu}$ and $\delta h = 1.1 L
^{-\gamma/\nu}$, respectively. Results for three different system
sizes are shown, namely for $L = 30$, $60$, and $L = 90$.

We observe again strong violations of self--averaging, obeying
finite--size scaling. They are not finite volume artifacts. We
have no theoretical explanation of these maximal violations of
self--averaging.

 The reader might be puzzled by our findings of a
bimodal distribution of $dm$ and at the same time the finite--size
scaling relation $\frac{\mathrm{d} m} {\delta h }  \sim L^{\gamma/
\nu } $. This is possible, as it is illustrated by the toy model
of the following bimodal probability distribution
\begin{equation}
\label{eq:6} p_{\rm toy} (dm, \delta h ,L )= p_0 \delta (dm) + p_1
\delta (dm-c ) \, ;\quad \ p_1 \sim  \delta h L^{\gamma/ \nu } \,
;\quad \ p_0 +p_1 = 1,
\end{equation}
where $c$ is a constant and $ \delta h $ is small. In this model
$\langle (dm)^k \rangle \sim \delta h c^k L^{\gamma/\nu}$.

\section{Conclusions}

In summary, numerical simulations have provided a significant step
forward in our understanding of the RFIM. The combination of an
appropriate fluctuation--dissipation formalism with modern
finite-size analysis~\cite{FMMa,FMMb} has provided strong evidence
for universality at space dimensions $d = 3$~\cite{FMMa,FMMb,PiS},
$d = 4$~\cite{FMMPSa,FMMPSa2}, and $d = 5$~\cite{FMMPSb}. The
evidence for violations of basic predictions of the PRG, such as
dimensional reduction, is crystal clear at $d = 3$ and $d = 4$. On
the other hand, our rather accurate results at $d = 5$ are
compatible with dimensional reduction. Although universality is a
prediction of the PRG, it is obvious that universality is valid
outside of the perturbative regime. An argument based on
eigenvalue-repulsion has been proposed to explain why universality
is robust against non--perturbative effects~\cite{Sourlas}.

Nevertheless many open questions remain. Our results are based on
numerical simulations with their inherent error bars. Even when
error bars are very small they do not replace a mathematical
proof. In particular we cannot exclude the possibility that PRG is
broken for any $d < D_{\rm u} = 6$ (a non--analytic dependence in
$d - D_{\rm u}$, such as $\exp{[1/(d-D_{\rm u})]}$, cannot be
excluded). It could also in theory be possible that such
exponentially small contributions violate universality.

If bound states are responsible for the breaking of the PRG, one
should introduce the fields representing these bound states and
write the effective Hamiltonian in terms of these fields. We do
not know what these fields are. This effective Hamiltonian can be
very different than the original one. The relation between the
fields of this Hamiltonian and the original ones can be very
complicated. A well known example in this context is the
sine--Gordon model where the effective field theory is the massive
Thirring model~\cite{mandelstam,coleman,mandelstam2}. The fermions
of the massive Thirring model are exponential functions of the
bosons of the sine--Gordon model.

\acknowledgments{We would like to thank Giorgio Parisi for his
hospitality in Rome, where part of this work has been completed.
V.M.-M. was partially supported by MINECO (Spain) through Grant
No. FIS2015-65078- C2-1-P (this contract partially funded by
FEDER). N.G.F. and M.P. acknowledge support by the Royal Society's
International Exchange Scheme 2016/R1.}

\end{document}